\documentclass[useAMS,usenatbib,usegraphicx]{mn2e}
\usepackage{amsmath}

\title[High energy $\gamma$-ray properties of the FR I radio galaxy NGC 1275]{High energy $\gamma$-ray properties of the FR I radio galaxy NGC 1275}

\author[Anthony M Brown]{Anthony M. Brown$^{1}$\thanks{E-mail:
anthony.brown@canterbury.ac.nz}\thanks{Corresponding author} and Jenni Adams$^{1}$\\
$^{1}$Department of Physics and Astronomy, University of Canterbury, Christchurch, 8140, New Zealand}
\begin{document}

\date{Accepted 2011 January 13. Received 2011 January 12. In original form 2010 December 15}

\pagerange{\pageref{firstpage}--\pageref{lastpage}} \pubyear{2011}

\maketitle

\label{firstpage}

\begin{abstract}

We report on our study of the high-energy $\gamma-$ray emission from the FR I radio galaxy NGC 1275, based on two years of observations with the \textit{Fermi}-LAT detector. Previous \textit{Fermi} studies of NGC 1275 had found evidence for spectral and flux variability on monthly timescales during the first year of \textit{Fermi}-LAT observations. This variability is also seen in the larger two year data set, during which we observe a large $\gamma-$ray flare (June-August 2010). The increased photon statistics from this large flare have allowed the discovery of flux variability from NGC 1275 on the timescales of days. The largest flux variation observed during this flare being a factor of $\sim 3$ from one day to the next and a resultant $e$-folding risetime of $1.51\pm0.2$ days. The two year averaged $E>$100 MeV $\gamma-$ray spectrum is adequately described by a power-law spectrum, with a photon index, $\Gamma$, of $2.09 \pm 0.02$, and a resultant integrated flux of $F_{\gamma}=(2.2\pm0.1) \times 10^{-7}$ ph cm$^{-2}$s$^{-1}$. While no hysteresis was observed in the photon index$-$flux ($\Gamma_{\gamma}$ vs F$_{\gamma}$) parameter space, there was obvious `harder-when-brighter' behaviour observed during the large $\gamma-$ray flare. Furthermore, during this large flare, NGC 1275 appeared to migrate from the FR I radio galaxy to the BL Lac object region of the photon index$-$luminosity ($\Gamma_{\gamma}$ vs L$_{\gamma}$) paramater space. In this paper we present details of our \textit{Fermi}-LAT analysis of NGC 1275, including a brief discussion on its implications for $\gamma-$ray blazar sources.

\end{abstract}


\section{Introduction}
NGC 1275, also known as Perseus A and 3C 84, is a nearby active galaxy located at the center of the Perseus cluster, at a redshift of z=0.0179 (\citet{red}). Classified as a FR I radio galaxy, NGC 1275 is extremely radio bright, exhibiting clear structure of a compact central source and an extended jet (e.g. \citet{radio1}, \citet{radio2}). VLBI and VOSP observations have shown NGC 1275's jets to have an instrinsic velocity of $0.5 \pm 0.09 c$ and an orientation angle $\approx 30-55\ensuremath{^{\circ}}$ (\citet{walker} \& \citet{asada2}). With a central supermassive black hole mass of $3.4 \times 10^8 \text{ M}_{\sun}$ (\citet{wilman}), NGC 1275 appears to exhibit jet precession (\citet{dunn}), which has been interpreted as a possible indication that NGC 1275 is the result of a merger between two galaxies (\citet{liu}). As a FR I radio galaxy, NGC 1275 is believed to belong to the parent population of BL Lac objects within the unified model of AGN (\citet{frref} and \citet{urry95}), though there is evidence that the AGN unification scheme may be over simplified (e.g., \citet{kharb}).

Surrounded by extended filament structures, NGC 2175 has historically been of great interest due to both its central location in the Perseus Cluster, with the Perseus cluster being the brightest cluster at X-ray energies (eg. \cite{pers_xray}), and the possible `feed-back' role NGC 1275 plays in the Perseus cluster (eg. \citet{gallagher}). Evidence for the `feed-back' role of NGC 1275 can be seen in both \textit{ROSAT} and \textit{Chandra} observations where cavity, or bubble, structures are visible (\citet{rosat}  and \citet{chandra}). These bubbles appear to be spatially coincident with the radio structure extending out from the central AGN engine. 

NGC 1275 is also of interest due to its close proximity to Earth affording us an excellent opportunity to study the physics of relativisitic outflows\footnote{Assuming $H_{0}=71$ km s$^{-1}$ Mpc$^{-1}$, NGC 1275 has a luminosity distance of 75.6 Mpc. At this distance 1 arcsecond corresponds to 324 parsecs.}. NGC 1275 was first detected as a source of high energy $\gamma$-rays by the \textit{Fermi}-LAT detector (\citet{firstfermi}). During the first four months of \textit{all-sky-survey} observations, taken from August to December 2008, NGC 1275 was found to have an average flux of $F_{\gamma} = (2.10 \pm 0.23) \times 10^{-7}$ ph cm$^{-2}$ s$^{-1}$ above 100 MeV and a power law photon index of $\Gamma = 2.17 \pm 0.05$. While no variability was observed during these four months of observations, the \textit{Fermi} detection of a $\gamma-$ray flux above the upper limit of $F_{>100 MeV} < 3.72 \times 10^{-8}$ ph cm$^{-2}$ s$^{-1}$ set by the \textit{EGRET} detector implied variability over larger timescales of years or decades (\citet{egret}).

More recently, Kataoka et al. have found evidence for monthly variations in the $E>100$ MeV $\gamma-$ray flux from NGC 1275 during the first year of \textit{Fermi}-LAT observations. These observations also showed spectral variations during periods of higher flux with a hysteresis behaviour in the flux-photon index parameter space (\citet{kataoka}). It is worth noting that during the first year of operations, NGC 1275 was continually present in the \textit{Fermi} bright source catalogue (\citet{bright1} \& \citet{bright2}).

At TeV $\gamma$-ray energies, upper limits for NGC 1275 above 100 GeV have been set by both the VERITAS Cherenkov telescope array (\citet{veritas}) and the MAGIC Cherenkov telescope array (\citet{magic}). These upper limits showed a deviation from extrapolating the power law spectrum at MeV/GeV energies down to TeV energies. However during the preparation of this paper, the MAGIC collaboration has recently announced the detection of NGC 1275 above 100 GeV with a significance of 5.2$\sigma$ at approximately 1\% of the Crab flux at TeV energies (\citet{vhe}).

This paper investigates the high-energy $\gamma-$ray flux and spectral properties of NGC 1275 during the first two years of \textit{Fermi}-LAT observations. In particular, the increased photon statistics associated with a large $\gamma$-ray flare event, allow us to probe flux and spectral variability from NGC 1275 with unprecedented temporal resolution. In \textsection 2 we describe the \textit{Fermi}-LAT observations and data analysis routines used in this study. The results of the analysis are given in \textsection 3, with the discussion in the context of AGN unification and TeV $\gamma$-ray emission, given in \textsection 4. The conclusions are given in \textsection 5. 

\section{\textit{Fermi}-LAT observations}
The LAT detector aboard \textit{Fermi}, described in detail in \citet{lat}, is a pair-conversion telescope, covering a photon energy range from below 20 MeV to above 300 GeV. With a large field of view, $ \simeq 2.4 $ sr, improved angular resolution, $\sim 0.6$\ensuremath{^{\circ}} at 1 GeV, and large effective area, $\sim 8000$ cm$^{2}$ on axis at 1 GeV, \textit{Fermi}-LAT provides an order of magnitude improvement in performance compared to previous $\gamma$-ray missions. 

Since the $4^{\text{th}}$ of August 2008, the vast majority of observations completed by \textit{Fermi} have been performed in \textit{all-sky-survey} mode. In \textit{all-sky-survey} mode \textit{Fermi} points away from the Earth and rocks $\sim32$\ensuremath{^{\circ}} north and south of its orbital plane. This rocking motion, coupled with \textit{Fermi}-LAT's large effective area, allows \textit{Fermi} to scan the entire $\gamma$-ray sky every 3 hours (\citet{ritz}). 

The observations used here comprise of all \textit{all-sky-survey} data taken between 4 August 2008 and 30 September 2010, equating to a MET interval of 239557417 to 307553159. After applying a zenith cut of 105\ensuremath{^{\circ}} to eliminate $\gamma$-ray photons from the Earth's limb, all `diffuse' class events were considered in the 100 MeV to 200 GeV energy range. Throughout this analysis, Science Tools version v9r15p2 were used in conjunction with instrument response functions (IRF) P6\_V3. Furthermore, a $\Lambda$CDM cosmology was adopted, with a Hubble constant of $H_{0}=71$ km s$^{-1}$ Mpc$^{-1}$, $\Omega_{m}=0.27$ and $\Omega_{\Lambda}=0.73$.

\section{Results}

\subsection{Sky Map} \label{sky}
\begin{figure}
 \centering
\includegraphics[width=80mm]{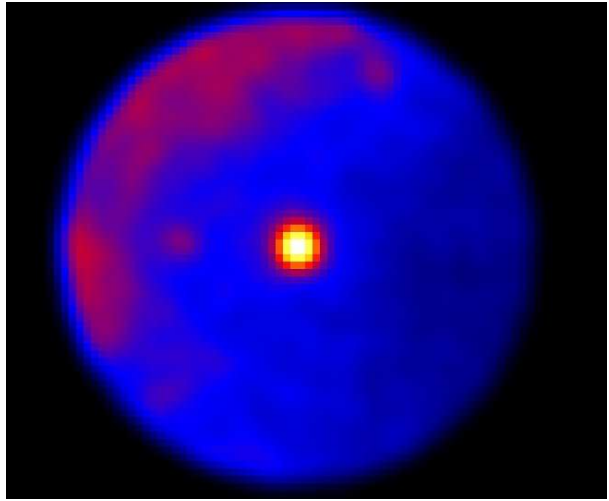}
\caption{100 MeV $-$ 200 GeV $\gamma$-ray image centered on NGC 1275 (RA $=$ 49.951\ensuremath{^{\circ}}, DEC$=$41.512\ensuremath{^{\circ}}) with a ROI of 8\ensuremath{^{\circ}}, smoothed with a 3\ensuremath{^{\circ}} Gaussian.}
\label{skymap}
\end{figure}

The 100 MeV $< E <$ 200 GeV sky map of NGC 1275 can be seen in Figure \ref{skymap}. Centered on NGC 1275 (RA$=$49.951\ensuremath{^{\circ}}, DEC$=$41.512\ensuremath{^{\circ}}), with a radius of interest (ROI) of 8\ensuremath{^{\circ}}, Figure \ref{skymap} has also been smoothed with a 3\ensuremath{^{\circ}} Gaussian function. The brightest source in this field of view is located at RA$=$49.951\ensuremath{^{\circ}}, DEC$=$41.533\ensuremath{^{\circ}}, and as such, is positionally coincident with NGC 1275. It should be noted that the extended structure towards the upper left of the field of view is believed to be diffuse $\gamma$-ray emission from the Galactic plane (\citet{firstfermi}). 

\subsection{Lightcurve}

\begin{figure*}
 \centering
 \begin{minipage}{180mm}
\includegraphics[width=120mm,angle=270]{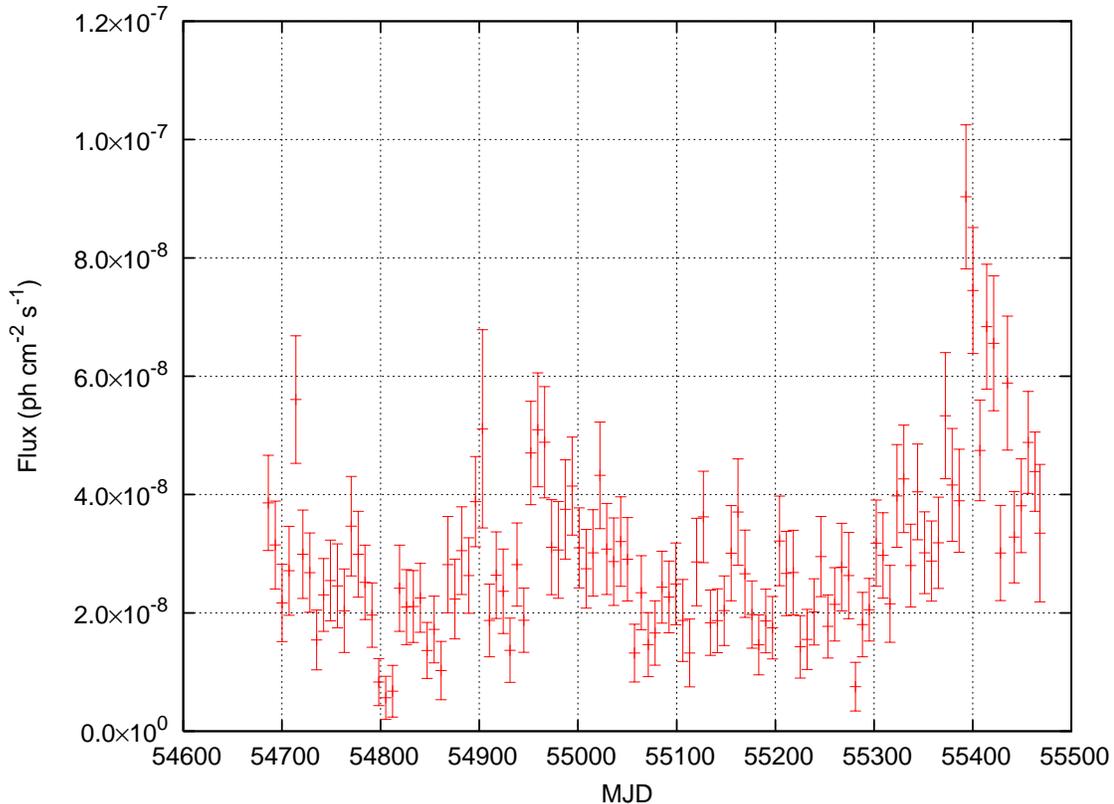}
\caption{The 2 year lightcurve of 800 MeV - 200 GeV $\gamma$-ray flux from NGC 1275 binned in week intervals, as detected by {\it Fermi}-LAT with a ROI of 1\ensuremath{^{\circ}}.}
\label{lc1}
\end{minipage}
\end{figure*}

To investigate temporal variability in NGC 1275's $\gamma$-ray flux, an unbinned maximum likelihood estimator, GTLIKE was applied to all 800 MeV $< E <$ 200 GeV diffuse class events, taking a ROI of 1\ensuremath{^{\circ}} to minimise the contribution of soft $\gamma$-rays from the Galactic ridge (see Section \ref{sky})\footnote{It should be noted that the lower limit on the ROI size is set by the angular resolution performance of the LAT instrument. At 800 MeV, 68\% of events are reconstructed within 1\ensuremath{^{\circ}} (\citet{lat}).}. Likewise, only considering events with $E > 800$ MeV allows us to remove the majority of the contribution from the Galactic diffuse emission. The P6\_V3\_DIFFUSE IRF was applied, assuming a power law spectrum, with an index of $\Gamma = 2.17$, as described in \cite{firstfermi}. The resultant lightcurve can be seen in Figure \ref{lc1} with a temporal resolution of a week and statistical errors only. Figure \ref{lc1} shows clear evidence for variability on timescales of weeks, with a $\chi^2 = 386$ for 112 d.o.f for a constant flux fit. Figure \ref{lc1} also reveals that several major flaring events have occurred during the two year observing period. The first flare during March to June 2009 (54850 $<$ MJD $<$ 54950) has previously been discussed in \citet{kataoka}. However, a larger flaring event is visible at the end of the observing period June to September 2010 (MJD $>$ 55350), with a maximum weekly flux of $(9.03 \pm 1.22) \times 10^{-8} \text{ ph cm}^{-2} \text{s}^{-1}$, increasing by a factor of 2.3 from the previous week.

\subsection{Spectrum}

To study NGC 1275's average $\gamma$-ray spectrum during the two years of observation GTLIKE was applied to the 100 MeV $< E <$ 200 GeV continuum, assuming a power law source model with additional Galactic and Extragalactic diffuse background components. Taking a ROI of 8\ensuremath{^{\circ}}, we utilised the most recent Galactic (gll\_iem\_v02.fit) and Extragalactic diffuse (isotropic\_iem\_v02.txt) models\footnote{It should be noted that the isotropic\_iem\_v02.txt model file also models any residual instrumental background.}, with the normalisation of these models free to vary. 

We obtain the following best-fit power law function for NGC 1275:
\begin{equation}
 \dfrac{dN}{dE}= (2.43 \pm 0.09) \times 10^{-9} (\dfrac{E}{100\text{ MeV}})^{-2.09\pm0.02} \nonumber
\end{equation}
\begin{equation}
 \text{ photons cm}^{-2} \text{s}^{-1} \text{MeV}^{-1}
\end{equation}
which equates to an integrated flux, in the 100 MeV $-$ 200 GeV energy range, of
\begin{equation}
  F_{E>100\text{ MeV}} = (2.2 \pm 0.1) \times 10^{-7}  \text{ photons cm}^{-2} \text{s}^{-1}
\end{equation}
only taking statistical errors into account. From this power law, the predicted $\gamma$-ray count was 6844.1, with a test statistic\footnote{The test statistic, TS, is defined as twice the difference between the log-likelihood of two different models ($2[\text{log} L - \text{log} L_{0}]$ where $L$ and $L_{0}$ are defined as the likelihood when the source is included or not respectively) \citet{mattox2}.} of $TS=9459.9$, corresponding to a $\sim 97\sigma$ detection. The Galactic diffuse background was found to have a normalisation of $1.07 \pm 0.01$ and a predicted count of 94496, while the Extragalactic diffuse background was found to have a normalisation of $0.77 \pm 0.02$ and a predicted count of 22118.5. The predicted diffuse background counts calculated from the two year data set are consistent with these previous works. However the photon index of NGC 1275 calculated over the two year data set, $\Gamma=(2.09\pm0.02)$, appears to be slightly harder those previously published in \cite{firstfermi}, $\Gamma=(2.17\pm0.04)$, from four months of observations and $\Gamma=(2.13\pm0.02)$, from one year of observations (\citet{kataoka}).

\begin{figure}
 \centering
\includegraphics[width=100mm,angle=270]{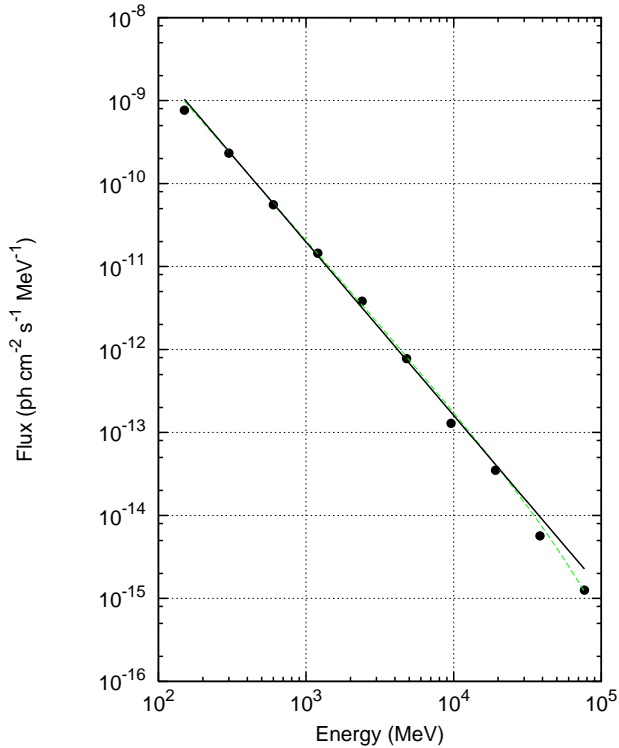}
\caption{The 2 year averaged LAT spectrum, $E>100$ MeV, of NGC 1275. The solid line represents the power law function defined in Equation 1, with the dashed line representing the exponential cut-off power law fit. With similar $\chi^2$ fits, it is not possible to differentiate between the two models.}
\label{spect}
\end{figure}

The best-fit power law function, given by Equation 1, can be seen in Figure \ref{spect}, where the data points in Figure \ref{spect} was obtained by applying GTLIKE separately to ten logarithmic energy bands in the range 100 MeV to 102.4 GeV. In an attempt to account for the small deviation between the model and data above $E=20$ GeV, the 100 MeV $-$ 102.4 GeV spectrum was also fit with an exponential cut-off power law ($dN/dE \propto E^{-\Gamma} \times \text{exp}(-E/E_{cut-off})$), with the best-fit shown in Figure \ref{spect}. However, it should be noted that the $\chi^2$ and resultant fit probability is similiar for both the power law and exponential cut-off power law fit. As such, it is not possible to differentiate between the two models for the averaged two year spectrum.

\section{Discussion}

\subsection{Flux Variability}
\begin{figure}
 \centering
\includegraphics[width=60mm,angle=270]{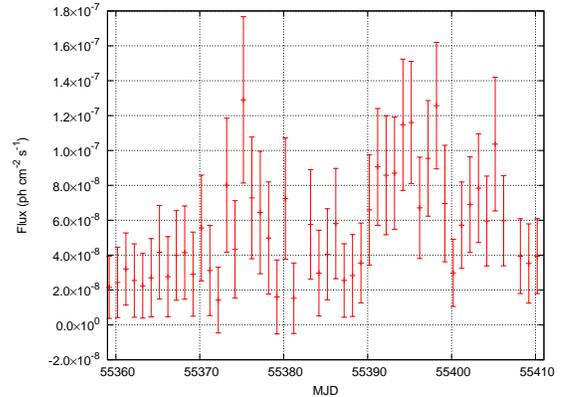}
\caption{Lightcurve of 800 MeV - 200 GeV $\gamma$-ray photons from NGC 1275 binned with daily temporal resolution, from June to August 2010, as detected by {\it Fermi}-LAT with a ROI of 1\ensuremath{^{\circ}}.}
\label{daylc}
\end{figure}

To investigate the major flare further, we rebinned the 800 MeV $< E <$ 200 GeV flux for June$-$August 2010 with daily resolution, the results of which can be seen in Figure \ref{daylc}. The most rapid flux variation is seen at 55373 $<$ MJD $<$ 55376  where the flux increased by a factor of $\sim$8 over a 3 day period. 

To characterise the timescales of this flare event, we evalulate the time for an exponential flux increase or decrease, referred to as the $e-$folding time, which is defined by:
\begin{equation}
 F(t)=F(t_{o})\text{exp}(\tau^{-1}(t-t_{o}))
\end{equation}
where $\tau$ is the characteristic $e-$folding timescale and $F(t)$ and $F(t_{o})$ are the fluxes at time $t$ and $t_{o}$ respectively. The variability observed during $55372<MJD<55380$ equates to an $e-$folding rise time, $\tau_{rise}$, of $1.51\pm0.2$ days and a subsequent $e-$folding decay time, $\tau_{decay}$, of $2.54\pm0.31$ days. 

With observed flux variability on timescales of days, it is most likely that the observed $\gamma$-ray variability from NGC 1275 is associated with relativistic outflows in the form of jets. This implies that relativistic effects such as Doppler boosting of radiation density and temporal contraction of variability timescales will be observed. Taking the Doppler factor of the relativistic jet into consideration, causality implies that an emission region, with radius, R, and a Doppler factor\footnote{$\delta = (\Gamma(1-\beta\text{cos}\theta))^{-1}$ where $\Gamma$ is the bulk Lorentz factor of the jet, $\beta=v/c$ and $\theta$ is the angle to the line of sight.}, $\delta$, is related to the $\gamma-$ray variability, $t_{var}$, by:
\begin{equation}
 R \leq ct_{var}\delta(1+z)^{-1}
\end{equation}

Taking $t_{var}$ to be the $e$-folding rise time, $\tau_{rise} = 1.51\pm0.2$ days constrains the emission region to $R\delta^{-1} \leq 3.84 \times 10^{13}$ meters.

By convolving this limit with that invoked by the optical depth of the $\gamma$-rays, $\tau_{\gamma\gamma} \leq 1$, we are able to estimate the minimum Doppler factor, $\delta_{min}$ (\citet{dondi} \& \citet{mattox}). Since no multiwavelength observations were taken during the flare period, we assume archival X-ray flux and spectrum values: $\alpha=0.65$ and $F_{5keV}=4$ $\mu$Jy (\citet{chu}). It should be noted that \textit{RXTE}-ASM data was used to confirm that the $2-10$ keV X-ray flux level during the Churazov et al. observations was similar to the $2-10$ keV flux level during the flare period analysed here. 

For $t_{var}=1.51$ days and 10 GeV photons we obtain a minimum Doppler factor of $\delta_{min} \approx 2$. A Doppler factor of 2 implies a maximum emission region radius of $\sim 8 \times 10^{13}$m or $2 \times 10^{-3}$ pc, strongly suggesting that the inner region of NGC 1275 is the source of the high-energy $\gamma$-ray emission.

We note that a Doppler factor of 2 is consistent with those derived for NGC 1275 via independent methods (eg. \citet{lah}), as well as being consistent with the Doppler factor derived for other \textit{Fermi} detected radio galaxies such as Cen A and M87 (\citet{cena}, \citet{m87}). It is worth mentioning that a Doppler factor of 2 is also consistent with broad-band modelling of Intermediate frequency BL Lac objects (IBL), which are believed to be the beamed counterparts of FR I radio galaxies within the unified model of AGN (eg. \citet{reyes}).

\subsection{Spectral Variability}

\begin{figure}
 \centering
\includegraphics[width=90mm]{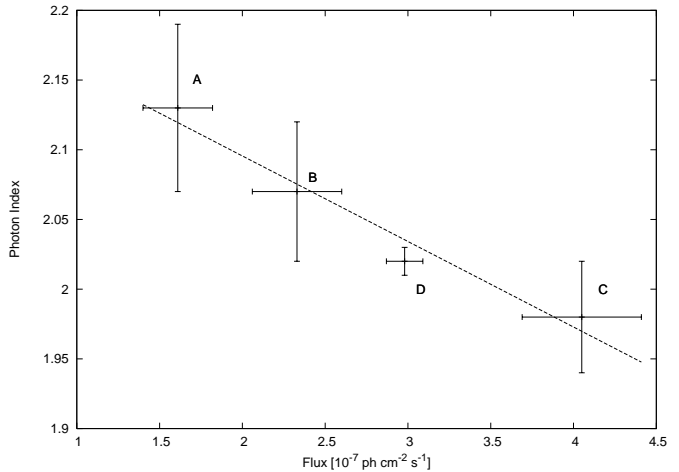}
\caption{Flux versus photon index parameter space for the four epochs `A', `B', `C' \& `D', assuming a power law spectrum, for the large flare MJD $>55200$.}
\label{fvsin}
\end{figure}

 Both \citet{firstfermi} and \citet{kataoka} found evidence for spectral variability in the $E>100$ MeV spectrum of NGC 1275. Furthermore, \cite{kataoka} found evidence for hysteresis behaviour in the flux versus photon index parameter space during the April$-$May 2009 flare. To investigate whether this hysteresis behaviour is present in the larger flare observed during June$-$August 2010, the MJD$>$55200 data was binned into four separate epochs corresponding to periods before, during the rise, at the peak and the decay of the flare event. These epochs, denoted as A, B, C and D, correspond to the MJD intervals (55200$<$MJD$<$55300), (55300$<$MJD$<$55372), (55372$<$MJD$<$55421) and (55421$<$MJD$<$55468) respectively. The unbinned maximum likelihood estimator was applied to each epoch, assuming a power law and the same diffuse emission models as before. The photon index and subsequent flux levels for epoch A, B, C and D can be seen in Figure \ref{fvsin}. 

While there is no evidence for hysteresis during the larger July$-$August 2010 flare, the flux level is clearly correlated with the photon index, such that the photon index becomes harder at higher flux levels ($ \Gamma \propto -0.06 \times F_{E>100 MeV} $). This `harder-when-brighter' behaviour is what one would expect in a simple acceleration and cooling scenario (\citet{kirk}) and has been observed previously with \textit{Fermi} observations of other AGN; eg. PKS 1502$+$106 and PKS 1510-089 (\citet{abdo_a} \& \citet{abdo_d}).

Morphological studies of \textit{Fermi} detected blazars have found the photon index to harden from 2.46 for Flat Spectrum Radio Quasars to 2.13 for IBL to 1.86 for High frequency peaked BL Lac objects (HBL) (\citet{abdo_b}). Furthermore, a study of the $E>100$ MeV properties of TeV-bright AGN has found that the majority of TeV AGN exhibit hard spectrum, with $\Gamma < 2$ (\citet{abdo_09c}). At its peak flux, NGC 1275 had a photon index of $\Gamma=1.98 \pm 0.04$, hardening from the $\Gamma=2.13 \pm 0.06$ observed during its `quiescent' state. This would imply that during the flare event, NGC 1275 migrated from a misaligned IBL to a HBL class of AGN. This migration has some interesting implications for detecting NGC 1275 at higher energies. \cite{kataoka} suggested that it is the $\geq GeV$ flux that is important when triggering TeV observations of \textit{Fermi}-LAT sources. This suggestion appears to be a valid one given that during NGC 1275's bright flare, when the spectrum was hardest, that is, when the GeV $\gamma$-ray flux was at its greatest, TeV $\gamma$-ray emission was detected by the MAGIC Cherenkov telescope array (\citet{vhe}). 

\subsection{High energy properties of FRI radio galaxies}

Now that three near FR I radio galaxies, Cen A, M87 and NGC 1275, have been detected by both \textit{Fermi}-LAT and ground based Cherenkov experiments, some preliminary comparisons can be made (see Table \ref{comparison}). 

While the MeV/GeV flux for NGC 1275 and Cen A are similiar, the photon indices are quite different ($\Gamma=1.98 \pm 0.04$ compared to $\Gamma=2.75 \pm 0.04$) (\citet{magn}). Meanwhile, M87 has a MeV/GeV flux that is an order of magnitude smaller than both NGC 1275 and Cen A, but a photon index that is harder than Cen A while being softer than NGC 1275 (\citet{magn}). While there is no obvious correlation between these observational characteristics, it is interesting to note that while M87 and NGC 1275 are known variable sources, Cen A does not appear to vary at $E>100$ MeV.

All of these FR I radio galaxies have been predicted to be strong sources of MeV$-$TeV $\gamma$-rays by a variety of emission models. One such model attributes the high energy emission to pulsar-like particle acceleration in a rotating magnetosphere around the supermassive black hole and was successfully applied to the variable $\gamma$-ray flux of M87 (\citet{neronov} \& \citet{rieger}). In this model $L_{IC} \propto M_{BH}$ is expected, where $L_{IC}$ is the inverse Compton luminosity and $ M_{BH}$ is the supermassive black hole (SMBH) mass. For FR I radio galaxies, the inverse Compton emission spans the MeV$-$TeV energy range and as such, if the rotating magnetosphere model was applicable to this class of AGN, one would expect to see both the MeV/GeV and TeV $\gamma$-ray fluxes and luminosities to be correlated to the SMBH mass. While this does appear to hold for the flux at TeV energies, it does not for the MeV/GeV flux. This would imply that the MeV/GeV and TeV emission have origins in different physical processes for radio galaxies. 

Finally, it is worth noting that FR I radio galaxies occupy a distinct region in the $\Gamma_{\gamma} - L_{\gamma}$ parameter space compared to BL Lac objects (\citet{magn}). This divide is attributed to a difference in accretion rate (\citet{ghiselleni}). During the large $\gamma$-ray flare presented in this paper, NGC 1275 migrated from the FR I radio galaxy to the BL Lac region of the $\Gamma_{\gamma} - L_{\gamma}$ parameter space with $\Delta \Gamma \sim 0.2$ and $\Delta \text{log L} \sim 0.62$. If indeed, the difference in $\Gamma_{\gamma} - L_{\gamma}$ parameter space is due to accretion rate, then the NGC 1275's flaring events of June$-$August 2010 could possibly be associated with an increase in accretion, and associated increase in the matter ejected along the relativistic jets.

\begin{table}
   \caption{Summary of high energy characteristics for nearby FR I radio galaxies. The Cen A and M87 MeV/GeV characteristics are from \citet{magn}, with the NGC 1275 MeV/GeV characteristics taken during the June$-$August 2010 flare when TeV emission was detected. The MeV/GeV flux is in units of $\text{photons cm}^{-2} \text{ s}^{-1}$. The TeV flux value for NGC 1275 was calcuated as 1\% of the Crab flux at TeV energies, with the TeV flux units of $\text{cm}^{-2} \text{ s}^{-1}$.}
   \begin{center}
     \begin{tabular}{llll} \hline \hline
               & Cen A       & M87     & NGC 1275         \\ \hline
   redshift    & 0.0009  & 0.004 & 0.0179       \\ 
   SMBH Mass ($\text{M}_{\sun}$)& $5.5 \times 10^{7}$ & $3.2 \times 10^{9}$   & $3.4 \times 10^{8}$   \\
   Flux (MeV/GeV) & $2.14 \times 10^{-7}$ & $2.45 \times 10^{-8}$ & $4.1 \times 10^{-7}$  \\ 
   $\Gamma$ (MeV/GeV)  & 2.75 & 2.21 & 1.98        \\
   Flux (TeV)      & $2.45 \times 10^{-13}$ & $1.5 \times 10^{-12}$ & $\sim 1 \times 10^{-12}$ \\ 
   $\Gamma$ (TeV) & 2.7 & 2.6 & $-$  \\ 
   log $L_{\gamma}$ (ergs s$^{-1}$) & 41.13 & 41.67 & 44.62\\ \hline
    \end{tabular}
  \end{center}
  \label{comparison}
\end{table}

\section{Conclusions}
We have reported on our studies of NGC 1275 utilising two years of \textit{all-sky-survey} data from the \textit{Fermi}-LAT detector. These studies have found NGC 1275 to exhibit $E>100$ MeV flux variability on timescales of days, with $\tau_{rise} = 1.51\pm0.2$ days, during a large $\gamma$-ray flare in June-August 2010. This large flare also saw NGC 1275 exhibiting spectral evolution, with a `harder-when-brighter' trend. This large flare period co-incided with the MAGIC collaboration announcing the detection of TeV emission from NGC 1275, giving weight to the belief that it is the $\geq GeV$ flux that is important when triggering observations at very high energies. 

The LAT spectrum of NGC 1275 is best described by a power law, with $\Gamma = 2.09\pm0.02$, though a power law with exponential cut-off cannot be ruled out.  This photon index is slightly harder than those of \cite{kataoka} and \cite{firstfermi}, with the difference most likely attributed to the increase in activity in the full two year data set compared to the one year and four month data sets.

In general, the spectral variability of NGC 1275 observed by \textit{Fermi} raises interesting questions with regards to the unification model of AGN. During the large flare, the $\gamma$-ray spectrum evolved from IBL to HBL values for photon indices. Likewise, in the $\Gamma_{\gamma} - L_{\gamma}$ parameter space, NGC 1275 migrated from the FR I radio galaxy to the BL Lac object region. Taking into consideration these two changes, it is no surprise that NGC 1275 has subsequently be detected at TeV energies, however, it raises the question as to why the three near FR I radio galaxies, NGC 1275, M87 and Cen A, all detected at both MeV/GeV and TeV energies, have such different high energy properties.

\section*{Acknowledgments}

This work is supported by the Marsden Fund Council from New Zealand Government funding, administered by the Royal Society of New Zealand. The authors would like to thank the referee, Ralph Wijers, for his useful comments and suggestions. This work has made use of public \textit{Fermi} data and \textit{RXTE}-ASM data obtained from the High Energy Astrophysics Science Archive Research Center (HEASARC), provided by NASA’s Goddard Space Flight Center. This work has also made use of the NASA/IPAC Extragalactic Database (NED), which is operated by the Jet Propulsion Laboratory, Caltech, under contact with the National Aeronautics and Space Administration.

\label{lastpage}

\end{document}